\documentclass[aps, prl, preprint, superscriptaddress, showpacs]{revtex4-1}
\usepackage{graphics}
\usepackage[english]{babel}
\usepackage{epstopdf}
\usepackage{amsmath}
\usepackage{amsfonts}
\usepackage{amssymb}
\usepackage[latin1]{inputenc} %caratteri speciali
\usepackage[OT2,T1]{fontenc}

\begin{document}

% Use the \preprint command to place your local institutional report
% number in the upper righthand corner of the title page in preprint mode.
% Multiple \preprint commands are allowed.
% Use the 'preprintnumbers' class option to override journal defaults
% to display numbers if necessary
%\preprint{}

%Title of paper

% repeat the \author .. \affiliation  etc. as needed
% \email, \thanks, \homepage, \altaffiliation all apply to the current
% author. Explanatory text should go in the []'s, actual e-mail
% address or url should go in the {}'s for \email and \homepage.
% Please use the appropriate macro foreach each type of information

%%%% AUTHOR LIST

\author{S. Finizio}
%\homepage[]{Your web page}
%\thanks{}
%\altaffiliation{}
\email{simone.finizio@psi.ch}
\affiliation{Swiss Light Source, Paul Scherrer Institut, 5232 Villigen PSI, Switzerland}

\author{S. Wintz}
\affiliation{Swiss Light Source, Paul Scherrer Institut, 5232 Villigen PSI, Switzerland}
\affiliation{Institute of Ion Beam Physics and Materials Research, Helmholtz-Zentrum Dresden-Rossendorf, 01328 Dresden, Germany}

\author{K. Zeissler}
\affiliation{School of Physics and Astronomy, University of Leeds, Leeds LS2 9JT, United Kingdom}

\author{A. V. Sadovnikov}
\affiliation{Laboratory Metamaterials, Saratov State University, Saratov, 410012, Russia}
\affiliation{Kotel'nikov Institute of Radioengineering and Electronics, Russian Academy of Sciences, Moscow, 125009, Russia}

\author{S. Mayr}
\affiliation{Swiss Light Source, Paul Scherrer Institut, 5232 Villigen PSI, Switzerland}
\affiliation{Department of Materials, Laboratory for Mesoscopic Systems, ETH Z\"urich, 8093 Z\"urich, Switzerland}

\author{S. A. Nikitov}
\affiliation{Laboratory Metamaterials, Saratov State University, Saratov, 410012, Russia}
\affiliation{Kotel'nikov Institute of Radioengineering and Electronics, Russian Academy of Sciences, Moscow, 125009, Russia}

\author{C. H. Marrows}
\affiliation{School of Physics and Astronomy, University of Leeds, Leeds LS2 9JT, United Kingdom}

\author{J. Raabe}
\affiliation{Swiss Light Source, Paul Scherrer Institut, 5232 Villigen PSI, Switzerland}

%%%% END OF AUTHOR LIST

\title{High resolution dynamic imaging of the delay- and tilt-free motion of N\'eel domain walls in perpendicularly magnetized superlattices}

%\date{\today}

%\doublespacing

\begin{abstract}
We report on the time-resolved investigation of current- and field-induced domain wall motion in perpendicularly magnetized microwires exhibiting asymmetric exchange interaction by means of scanning transmission x-ray microscopy using a time step of 200 ps. Dynamical domain wall velocities on the order of 50-100 m s$^{-1}$ were observed. The improvement in the temporal resolution allowed us to observe the absence of incubation times for the motion of the domain wall, together with indications for a negligible inertia. Furthermore, we observed that, for short current and magnetic field pulses, the magnetic domain walls do not exhibit a tilting during its motion, providing a mechanism for the fast, tilt-free, motion of magnetic domain walls.
\end{abstract}

% insert suggested PACS numbers in braces on next line`
%\pacs{68.37.Yz, 75.80.+q, 75.78.Cd}
% 68.37.Yz - X-ray microscopy
% 75.80.+q - Magnetostriction
% 75.78.Cd - Micromagnetic simulations

% insert suggested keywords - APS authors don't need to do this
%\keywords{}

%\maketitle must follow title, authors, abstract, \pacs, and \keywords
\maketitle

Since the first observations that electrical currents can induce the motion of magnetic domain walls (current induced domain wall motion - CIDWM) through the spin transfer torque effect \cite{art:ralph_STT, art:hayashi_STT_CIDWM}, many different mechanisms that allow for the motion of magnetic domain walls by electrical currents have been discovered. In particular, the discovery that spin-orbit torques can displace magnetic domain walls \cite{art:gambardella_SOT_Review, art:miron_CIDWM, art:ryu_CIDWM, art:parkin_CIDWM, art:loconte_CIDWM, art:vogel_TR_CIDWM, art:taniguchi_CIDWM} has allowed for the achievement of fast domain wall motion velocities (several hundreds of meters per second) in multilayer superlattice stacks optimized for both a high perpendicular magnetic anisotropy (PMA) and a large asymmetric exchange interaction (or Dzyaloshinskii-Moriya - DM - interaction). These discoveries have allowed for the development and optimization of magnetic memory concepts based on CIDWM such as e.g. the racetrack memory \cite{art:parkin_CIDWM}. 

The unraveling of the domain wall motion dynamics is of paramount importance for the understanding of the physical processes behind their motion, such as e.g. the acceleration/deceleration of the domain wall at the start/end of the application of the electrical current (i.e. the inertia of the domain wall) \cite{art:vogel_TR_CIDWM, art:torrejon_DW_inertia}, the effect of the DM interaction on the shape of the domain wall during its motion \cite{art:boulle_FIDWM, art:jue_FIDWM}, and the differences between current- and field-induced domain wall motion (FIDWM) \cite{art:boulle_FIDWM, art:jue_FIDWM}. Furthermore, the understanding of the dynamical processes is required in order to validate the theoretical models (such as e.g. the 1D model \cite{art:ryu_CIDWM}) currently employed for the description of the CIDWM and FIDWM processes.

The experimental investigation of the sub-ns temporal dynamics of the CIDWM and FIDWM processes requires a non-invasive microscopy technique combining a high spatial and temporal resolution (in the pump-probe regime) with an insensitivity to electric and magnetic fields. Only a few time-resolved studies of the CIDWM process exist \cite{art:vogel_TR_CIDWM}, but they are limited in temporal resolution (in Ref. \cite{art:vogel_TR_CIDWM}, time steps on the order of 5-20 ns are employed) and in the magnitude of the applicable current densities (due to the use of photoemission electron microscopy, which is extremely sensitive to variations in the surface potential of the sample \cite{art:schoenhense_peem_probing_depth}, as the probing technique).

In this letter, we report on the time-resolved investigation of the CIDWM and FIDWM processes in a Pt/Co$_{68}$B$_{32}$/Ir microwire using time-resolved scanning transmission x-ray microscopy (STXM). Time-resolved images of the dynamics of the CIDWM and FIDWM processes employing a time step of 200 ps were successfully acquired, and we could observe an indication of negligible inertia of the domain wall, along with the absence of a tilt of the domain wall when applying short current/field pulses.

\begin{figure}
 \includegraphics{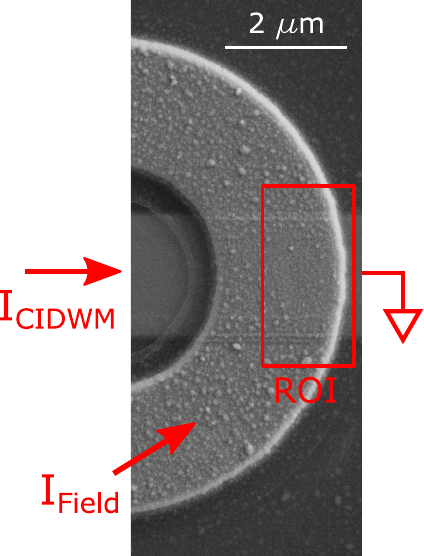}
 \caption{Scanning electron micrograph of the Pt/Co$_{68}$B$_{32}$/Ir microwire employed for the experiments presented here. A current pulse is injected from the left-hand electrode, with the right-hand electrode at the sample ground, and an out-of-plane magnetic field pulse is generated by injecting a current pulse across the $\Omega$-shaped microcoil fabricated on top of the microwire. The microwire and the $\Omega$-shaped coil are electrically insulated through a SiO$_2$ layer. The area investigated in the experiments reported here (region of interest - ROI) is marked by a red rectangle in the figure.}
 \label{fig:SEM}
\end{figure}

\begin{figure}
 \includegraphics{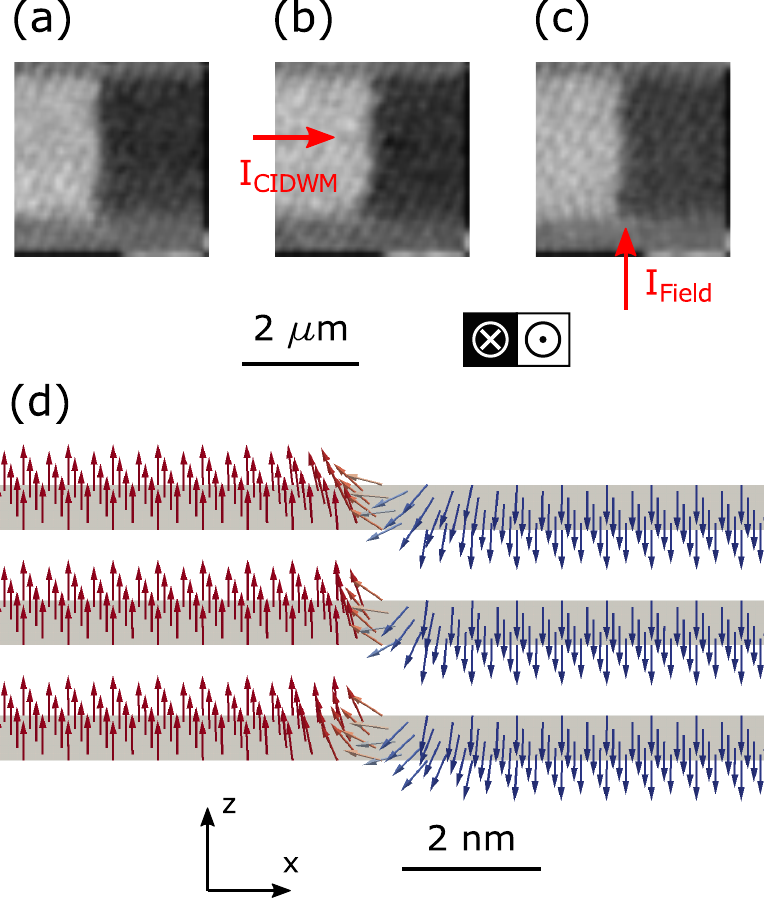}
 \caption{Quasi-static XMCD-STXM images of (a) the initial magnetic configuration, obtained by generating several magnetic field pulses with the $\Omega$-shaped coil; (b) motion of the magnetic domain wall caused by the injection of a 5 ns current pulse across the Pt/Co$_{68}$B$_{32}$/Ir microwire; (c) Restoration of the initial magnetic configuration by generating one magnetic field pulse with the $\Omega$-shaped coil. (d) Micromagnetic simulation of a domain wall in a three-layer Pt/Co$_{68}$B$_{32}$/Ir superlattice stack, where it is possible to observe that the domain wall is N\'eel-type, with the same chirality across all three layers.}
 \label{fig:static_set_reset}
\end{figure}

Microstructured thin film wires of 2 $\mu$m width were fabricated by electron beam lithography followed by liftoff out of a Ta(3.2 nm)/Pt(2.6 nm)/[Co$_{68}$B$_{32}$(0.8 nm)/Ir(0.4 nm)/Pt(0.6nm)]$_{\times 3}$/Pt(2.1 nm) multilayer superlattice stack deposited by sputtering (details on the deposition of a similar multilayer superlattice stack are given in Ref. \cite{art:zeissler_pinning}). The microwires were fabricated on top of an x-ray transparent Si$_3$N$_4$ membrane on a high resistivity Si frame. The microwires were contacted by 200 nm thick Cu electrodes fabricated by electron-beam lithography followed by a lift-off process. To generate the magnetic field pulses employed for the reset of the original magnetic state, an $\Omega$-shaped, 400 nm thick, Cu coil was fabricated on top of the microwire by electron-beam lithography followed by a lift-off process. The electrical insulation between the microwire and the $\Omega$-shaped coil is guaranteed by a 200 nm thick SiO$_2$ layer, deposited by electron-beam evaporation, between the microwire and the coil. A scanning electron micrograph of the sample is shown in Fig. \ref{fig:SEM}. The measurements of the CIDWM and FIDWM dynamics were carried out in a region of the Pt/Co$_{68}$B$_{32}$/Ir microwire where the current density is uniformly distributed across the microwire, as verified by finite element simulations of the current flow (see the area marked by ROI in Fig. \ref{fig:SEM}). Due to the contribution from the interfacial DM interaction \cite{art:zeissler_pinning}, the Pt/Co$_{68}$B$_{32}$/Ir multilayered superlattices stabilize N\'eel-type domain walls. Due to the low number of repeats, it is expected that the chirality of the N\'eel-type domain wall will be the same for all of the three layers \cite{art:legrand_bloch_walls}. We verified this statement by conducting micromagnetic simulations of a N\'eel-type domain wall stabilized in the Pt/Co$_{68}$B$_{32}$/Ir stack employing MuMax3 \cite{art:mumax}. For the simulations, a saturation magnetization of M$_\mathrm{s}$ = 665 kA m$^{-1}$, a PMA of k = 600 kJ m$^{-3}$, and a value of D = 1.17 mJ m$^{-2}$ for the DM interaction were selected. These values were determined by magnetometry and Brillouin Light Scattering measurements from a Pt/Co$_{68}$B$_{32}$/Ir film grown in parallel to the ones that were patterned. A simulation grid of 0.1$\times$0.1$\times$0.1 nm$^3$ was employed for the simulations. Three magnetic layers of 0.8 nm thickness separated by non-magnetic layers of 1 nm thickness were simulated, using the same set of boundary conditions employed in Ref. \cite{art:legrand_bloch_walls}. The result of the micromagnetic simulations is shown in Fig. \ref{fig:static_set_reset}(d), where it is possible to observe that a N\'eel-type domain wall of the same chirality is stabilized in all of the three layers of the Pt/Co$_{68}$B$_{32}$/Ir film.

The dynamics of the CIDWM and FIDWM processes were imaged using STXM at the PolLux (X07DA) endstation of the Swiss Light Source \cite{art:pollux}. A Fresnel zone plate with an outermost zone width of 25 nm was employed for the measurements presented here. The width of the entrance and exit slits to the monochromator was selected to achieve a spatial resolution on the order of 25 nm. Magnetic contrast was achieved by the x-ray magnetic circular dichroism (XMCD) effect \cite{art:schuetz_xmcd} by tuning the circularly-polarized x-rays to the Co L$_3$ absorption edge (about 778 eV). The quasi-static images (e.g. Fig. \ref{fig:static_set_reset}) were obtained by acquiring two images with opposite photon helicities, and calculating the XMCD contrast by subtracting the two images. Pump-probe time-resolved images were acquired employing circularly-polarized photons with negative helicity, with an avalanche photodiode detector combined with a dedicated field-programmable gate array setup \cite{art:finizio_TR_STXM, art:puzic_TR_STXM}. A time step of 200 ps was employed for the time-resolved measurements presented in this letter. The time position of the zero delay between pump and probe (also referred to as $t_0$) was determined with a fast laser diode connected to the same electronic setup employed for the excitation. The calibration of $t_0$ allowed for an accuracy of below 200 ps in the synchronization between the excitation signal and the recorded dynamical processes, providing a substantial improvement with respect to the first time-resolved measurements reported for CIDWM in PMA systems \cite{art:vogel_TR_CIDWM}.

To guarantee the reproducibility necessary for carrying out pump-probe measurements, the following experimental protocol, shown in Fig. \ref{fig:static_set_reset} with static XMCD-STXM images, was adopted: the sample was prepared in its original state by saturating it with a static magnetic field (of about 200 mT, well above the coercive field of the Pt/Co$_{68}$B$_{32}$/Ir multilayer), and ten thousand 4 ns-long pulses with a peak current of 350 mA were injected across the $\Omega$-shaped coil, leading to the magnetic state shown in Fig. \ref{fig:static_set_reset}(a). Following the initialization of the sample, the time-resolved measurements were carried out with the following procedure: (i) a 5 ns long current pulse, leading to a peak current density of $9 \cdot 10^{11}$ A m$^{-2}$, was injected across the microwire, leading to a current-induced displacement of the domain wall (Fig. \ref{fig:static_set_reset}(b)); (ii) a 4 ns-long, 350 mA, current pulse was injected across the $\Omega$-shaped coil, leading to the generation of a magnetic field that displaced the domain wall back to its original position (Fig. \ref{fig:static_set_reset}(c)). 

As expected from the contribution of the spin-orbit torques to the domain wall motion (caused by the spin Hall effect, leading to the generation of a transverse spin current in the heavy metal present in the multilayered superlattice stack \cite{art:gambardella_SOT_Review}), the magnetic domain walls are displaced along the direction of the electrical current (i.e. opposite to the direction of the electron flow) \cite{art:gambardella_SOT_Review, art:miron_CIDWM, art:ryu_CIDWM, art:parkin_CIDWM, art:loconte_CIDWM, art:vogel_TR_CIDWM}. For a 5 ns-long current pulse, a displacement of about 250 nm was observed for the magnetic domain wall, leading to an average domain wall velocity of 50 m s$^{-1}$, in agreement with the observed velocities for similar multilayer superlattice stacks \cite{art:vogel_TR_CIDWM, art:loconte_CIDWM, art:miron_CIDWM, art:taniguchi_CIDWM}. 

\begin{figure}
 \includegraphics{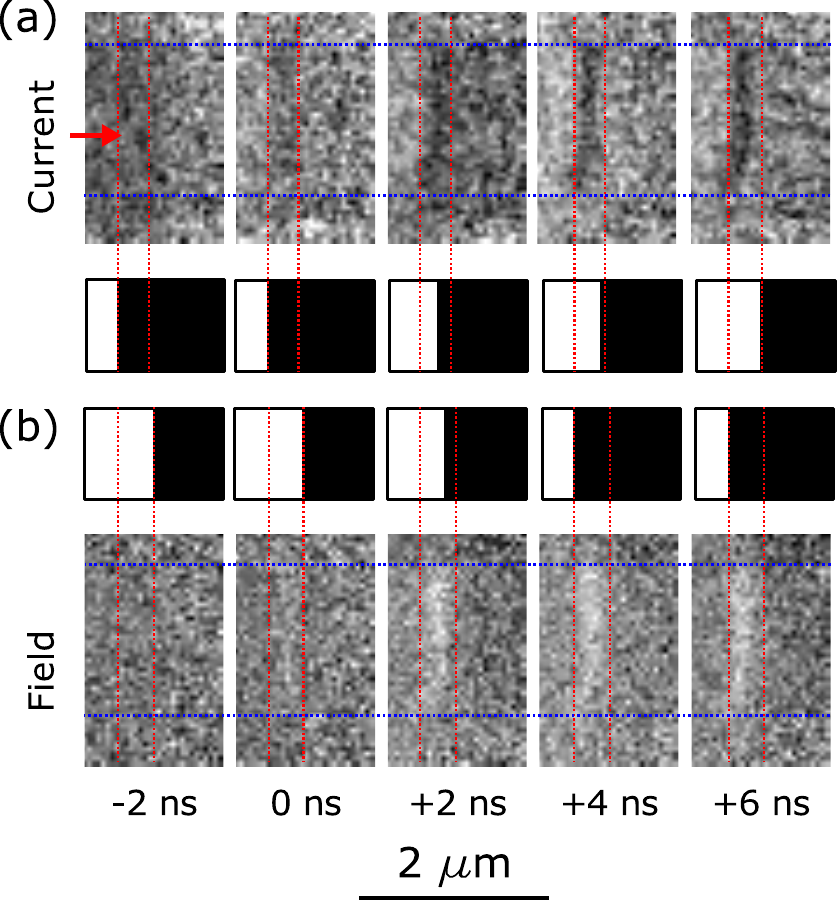}
 \caption{Snapshots at different time delays of (a) the CIDWM (the direction of the applied current is marked by a red arrow in the figure) and (b) the FIDWM processes. The snapshots are shown as differential images. The position of the domain wall is sketched next to each of the snapshots. The position of the microwire is marked by blue dashed lines in the figure.}
 \label{fig:dynamic_CIDWM}
\end{figure}

Figure \ref{fig:dynamic_CIDWM} shows snapshots of a time-resolved image of the CIDWM and FIDWM processes (2 ns time step). In order to better visualize the domain wall motion, the images shown in Fig. \ref{fig:dynamic_CIDWM} show the variation in the spin configuration at each frame with respect to the magnetic configuration after the restoration of the original magnetic configuration with the magnetic field pulse for (a) and with respect to the magnetic state after the CIDWM for (b). In contrast to XMCD-PEEM imaging, STXM is insensitive to the surface potential, therefore allowing us to investigate the domain wall motion at high current densities without the image distortions and losses of focus present for PEEM imaging of microwires under an applied voltage difference \cite{art:vogel_TR_CIDWM}.

\begin{figure}
 \includegraphics{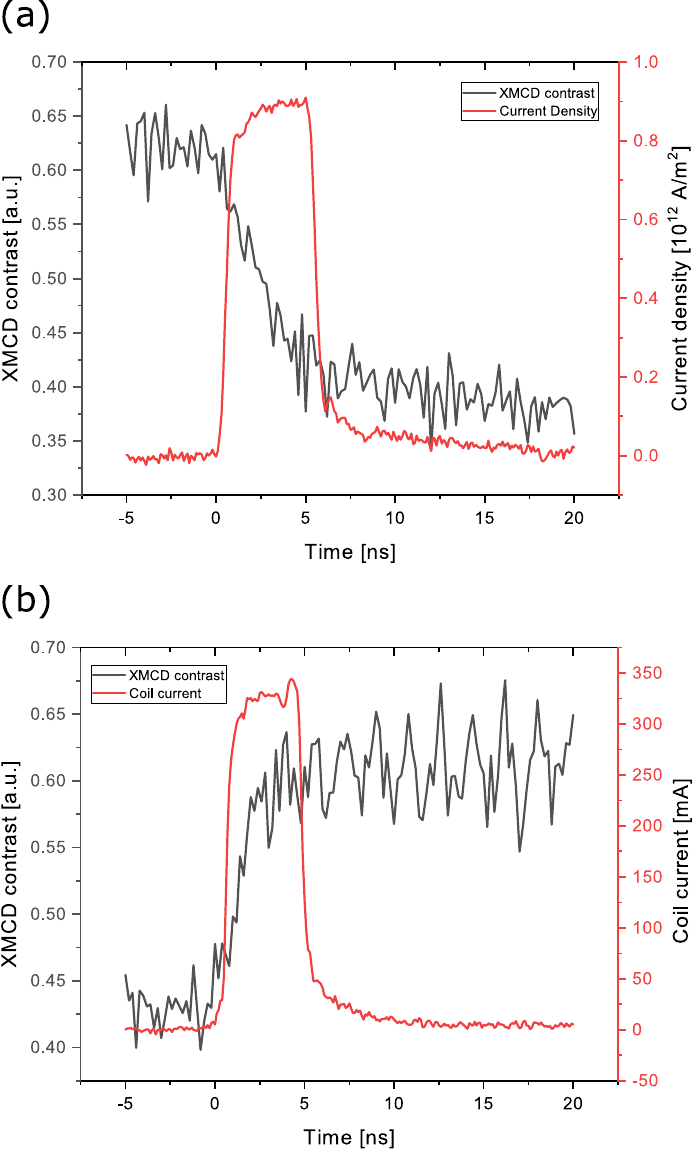}
 \caption{Time resolved XMCD contrast variation along the magnetic domain wall upon (a) the injection of a current across the microwire, and (b) upon the injection of a current across the $\Omega$-shaped coil. A delay-free motion of the magnetic domain wall can be observed for both the CIDWM and the FIDWM processes. From the time traces, a domain wall velocity of 50 m s$^{-1}$ and of 100 m s$^{-1}$ can be observed for the CIDWM and FIDWM processes respectively.}
 \label{fig:time_traces}
\end{figure}

By determining the changes in the magnetic contrast over the excitation cycle (for the time-resolved images with a 200 ps temporal resolution), shown in Fig. \ref{fig:time_traces} for both the CIDWM (\ref{fig:time_traces}(a)) and FIDWM (\ref{fig:time_traces}(b)) processes, additional information on the dynamics can be gathered. An immediately observable difference between the dynamics of the CIDWM and FIDWM processes is that the current-induced motion of the domain wall halts only at the end of the current pulse, while the field-induced motion halts about 2.4 ns after the start of the field pulse. This difference underlines a different average domain wall velocity for the CIDWM and FIDWM processes, of about 50 m s$^{-1}$ at a peak current density of $9 \cdot 10^{11}$ A m$^{-2}$ for the CIDWM, and of about 100 m s$^{-1}$ for the FIDWM at a peak magnetic field of about 70 mT. The absence of motion after 2.4 ns from the start of the magnetic field pulse can be explained if one considers the distribution of the magnetic field generated by the $\Omega$-shaped coil, shown in the inset of Fig. \ref{fig:time_traces}(b). The spatial distribution of the out-of-plane magnetic field generated by the $\Omega$-shaped coil provides a minimum in the energy for the position of the magnetic domain wall and, once the domain wall reaches that position, the motion will halt. It is worth to note here that the oscillations of the recorded contrast after the magnetic field pulse in Fig. \ref{fig:time_traces}(b) are a measurement artifact due to the cross talk between the $\Omega$-shaped coil and the avalanche photodiode detector of the STXM setup.

For both the CIDWM and FIDWM processes, the motion of the domain wall starts synchronously with the current or field excitation, at least within the available time step of 200 ps. The observation of the absence of an incubation time for the motion of the domain walls provides a strong confirmation that the N\'eel domain walls stabilized in the Pt/Co$_{68}$B$_{32}$/Ir PMA multilayer superlattice stack reported here exhibit a negligible inertia, similarly to previous observations on Pt/Co/AlO$_\mathrm{x}$ thin films \cite{art:vogel_TR_CIDWM}. This observation is further backed by the sudden halt of the domain wall motion at the end of the current pulse, indicating a fast deceleration of the domain wall (see Fig. \ref{fig:time_traces}(a)).

Furthermore, as shown in Fig. \ref{fig:dynamic_CIDWM}, the magnetic domains do not appear to tilt during their displacement induced either by the magnetic field pulse or by the electrical current, remaining therefore normal to the wire axis. This seems, at a first glance, in contrast with the models proposed in Refs. \cite{art:boulle_FIDWM, art:torrejon_DW_inertia}, and with quasi-static experimental observations \cite{art:garg_Y_CIDWM}. However, in agreement with the predictions reported in Ref. \cite{art:jue_FIDWM}, microwires with a large width (such as the 2 $\mu$m wires reported here) exhibit a long transient before reaching a steady-state motion with a defined domain wall tilt. At the start of such transient, fitting our experimental observations, the motion of the domain wall can be approximated as tilt-free. Therefore, a possible approach to reduce the effect of domain wall tilting while still keeping a reliable domain wall motion could be to employ short current/field pulses. The absence of an incubation time and the negligible inertia of these domain walls will guarantee that the short current/field pulses will displace the domain wall, whilst however being too fast to cause a sizable tilting of the domain wall. The application of a train of short current/field pulses could then be employed for long-distance displacement of the magnetic domain walls. This motion protocol is similar to the one reported in Ref. \cite{art:garg_Y_CIDWM}, where it was observed that short current pulses lead to a different behavior of the domain wall motion if compared to longer current pulses.

In conclusion, we have reported on the time-resolved imaging of the current- and field-induced domain wall dynamics in microwires fabricated out of Pt/Co$_{68}$B$_{32}$/Ir multilayered superlattice stacks exhibiting interfacial DM interaction at high spatial (25 nm) and temporal (sub-200 ps) resolution using time-resolved STXM. In particular, the domain wall motion under the application of short (5 ns) current and (4 ns) magnetic field pulses was investigated. For both the current- and field-induced motion, the absence of an incubation time (within the available time step of 200 ps) was observed, suggesting a low inertial mass for the N\'eel domain walls stabilized in the Pt/Co$_{68}$B$_{32}$/Ir films. Furthermore, the absence of a tilting of the N\'eel domain walls when displaced by the short current/field pulses was observed, providing a possible motion protocol enabling a long-range tilt-free motion of the domain walls.

The substantial improvement in the temporal and spatial resolution that we report here combined with the insensitivity of STXM imaging to the surface potential of the sample will allow for the investigation of a wide range of current densities and pulse durations. Future measurements with longer pulse durations can e.g. be aimed at the investigation of the domain wall tilting process as a function of the pulse duration, or of the acceleration and deceleration processes of the magnetic domain walls in larger wires \cite{art:torrejon_DW_inertia}.

%% If you have acknowledgments, this puts in the proper section head.

\begin{acknowledgments}
%SF conceived the experiment, with input from JR, SW, KZ, and CHM. SF and KZ designed the sample. KZ grew the Pt/Co$_{68}$B$_{32}$/Ir multilayer superlattice stacks. KZ, AVS, and SAN characterized the magnetic properties of the Pt/Co$_{68}$B$_{32}$/Ir multilayer superlattice stacks. SF lithographically patterned the samples. SF, SW, SM, and JR performed the time-resolved STXM experiments. SF, with SW and JR, analyzed the time-resolved data. SF wrote the manuscript, which was then discussed with all authors.

This work was performed at the PolLux (X07DA) endstation of the Swiss Light Source, Paul Scherrer Institut, Villigen, Switzerland. The research leading to these results has received funding from the European Community's Seventh Framework Programme (FP7/2007-2013) under grant agreement No. 290605 (PSI-FELLOW/COFUND), and the European Union's Horizon 2020 Project MAGicSky (Grant No. 665095). The authors thank P. Gambardella for helpful discussions.
\end{acknowledgments}

\bibliography{TR_CIDWM_bibliography}

\end{document}